\documentclass[aps,prd,twocolumn,amsfonts,amssymb,amsmath,showpacs,letterpaper,superscriptaddress,nofootinbib,altaffilletter,fontenc]{revtex4-2}
\usepackage[utf8]{inputenc}
\usepackage[T1]{fontenc}

\usepackage{graphicx}	% Including figure files
\graphicspath{ {./Figures/} }
\usepackage{caption}
\usepackage{subcaption}
\usepackage{amsmath}	% Advanced maths commands
\usepackage{amssymb}	% Extra maths symbols
\usepackage{tabularx}
\usepackage{enumerate}
\usepackage{enumitem}
\usepackage{comment}
\usepackage{multirow}
\usepackage{float}
\usepackage[citecolor=blue,linkcolor=blue,colorlinks=true,urlcolor=blue]{hyperref}
\usepackage[misc]{ifsym}
\usepackage{orcidlink}

\newcommand{\cqg}{Class. Quantum Grav.}

\begin{document}

\title{Gravitational Wave Detector Sensitivity to Eccentric Black Hole Mergers}
\begin{comment}
\author[Shubhagata Bhaumik et al.]{
Shubhagata Bhaumik,$^{1}$
V. Gayathri,$^{2}$ 
Imre Bartos,$^{1}$\thanks{E-mail: imrebartos@ufl.edu} 
Jeremiah Anglin,$^{1}$
Gregorio Carullo,$^{3,4}$
James Healy,$^{5}$
\newauthor
Sergey Klimenko,$^{1}$
Jacob Lange,$^{6}$
Carlos Lousto,$^{5}$
Tanmaya Mishra,$^{1}$
Marek J. Szczepa\'nczyk$^{7}$
\\
% List of institutions
$^{1}$Department of Physics, University of Florida, PO Box 118440, Gainesville, FL 32611-8440, USA\\
$^{2}$Leonard E. Parker Center for Gravitation, Cosmology, and Astrophysics, University of Wisconsin–Milwaukee, Milwaukee, WI 53201, USA\\
$^{3}$Niels Bohr International Academy, Niels Bohr Institute, Blegdamsvej 17, 2100 Copenhagen, Denmark\\
$^{4}$School of Physics and Astronomy and Institute for Gravitational Wave Astronomy, University of Birmingham, Edgbaston, Birmingham, B15 2TT, United Kingdom\\
$^{5}$Center for Computational Relativity and Gravitation, School of Mathematical Sciences, Rochester Institute of Technology, 85 Lomb Memorial Drive, \\Rochester, New York 14623, USA\\
$^{6}$Center for Gravitational Physics, The University of Texas at Austin, Austin, Texas 78712, USA\\
$^{7}$Faculty of Physics, University of Warsaw, Ludwika Pasteura 5, 02-093 Warsaw, Poland\\
%$^{4}$Department, Institution, Street Address, City Postal Code, Country
}
\end{comment}

\author{Shubhagata Bhaumik}
\email[E-mail: ]{sbhaumik@ufl.edu}
\affiliation{Department of Physics, University of Florida, PO Box 118440, Gainesville, FL 32611-8440, USA}

\author{V. Gayathri}
\affiliation{Leonard E. Parker Center for Gravitation, Cosmology, and Astrophysics, University of Wisconsin–Milwaukee, Milwaukee, WI 53201, USA}

\author{Imre Bartos}
\affiliation{Department of Physics, University of Florida, PO Box 118440, Gainesville, FL 32611-8440, USA}

\author{Jeremiah Anglin}
\affiliation{Department of Physics, University of Florida, PO Box 118440, Gainesville, FL 32611-8440, USA}

\author{Gregorio Carullo}
\affiliation{Niels Bohr International Academy, Niels Bohr Institute, Blegdamsvej 17, 2100 Copenhagen, Denmark}
\affiliation{School of Physics and Astronomy and Institute for Gravitational Wave Astronomy, University of Birmingham, Edgbaston, Birmingham, B15 2TT, United Kingdom}

\author{James Healy}
\affiliation{Center for Computational Relativity and Gravitation, School of Mathematical Sciences, Rochester Institute of Technology, 85 Lomb Memorial Drive, Rochester, New York 14623, USA}

\author{Sergey Klimenko}
\affiliation{Department of Physics, University of Florida, PO Box 118440, Gainesville, FL 32611-8440, USA}

\author{Jacob Lange}
\affiliation{Center for Gravitational Physics, The University of Texas at Austin, Austin, Texas 78712, USA}

\author{Carlos Lousto}
\affiliation{Center for Computational Relativity and Gravitation, School of Mathematical Sciences, Rochester Institute of Technology, 85 Lomb Memorial Drive, Rochester, New York 14623, USA}

\author{Tanmaya Mishra}
\affiliation{Department of Physics, University of Florida, PO Box 118440, Gainesville, FL 32611-8440, USA}

\author{Marek J. Szczepa\'nczyk}
\affiliation{Faculty of Physics, University of Warsaw, Ludwika Pasteura 5, 02-093 Warsaw, Poland}

%\date{}
%\pubyear{\the\year{}}

%\label{firstpage}

\begin{abstract}
Orbital eccentricity in compact binary mergers carries crucial information about the binary's formation and environment. There are emerging signs that some of the mergers detected by the LIGO and Virgo gravitational wave detectors could indeed be eccentric. Nevertheless, the identification of eccentricity via gravitational waves remains challenging, to a large extent because of the limited availability of eccentric gravitational waveforms. While multiple suites of eccentric waveforms have recently been developed, they each cover only a part of the binary parameter space. 
%In addition, there has been no systematic study so far to evaluate the reliability of these waveforms. 
Here we evaluate the sensitivity of LIGO to eccentric waveforms from the SXS and RIT numerical relativity catalogs and the TEOBResumS-Dali waveform model using data from LIGO-Virgo-Kagra's third observing run. The obtained sensitivities, as functions of eccentricity, mass and mass ratio, are important inputs to understanding detection prospects and observational population constrains. In addition, our results enable the comparison of the waveforms to establish their compatibility and applicability for searches and parameter estimation. 
%\vspace{1cm}
\end{abstract}

\date[\relax]{Dated: \today }

\maketitle

\begin{comment}
    \begin{keywords}
gravitational waves -- eccentric binary black holes -- dynamical formation
\end{keywords}
\end{comment}

\section{Introduction}

Gravitational waves from black hole binaries carry information about the origin and environment of the mergers, and the relevant astrophysical processes. Typically, the black holes' mass and spin is being tapped for this information \citep{Belczynski_2002, Rodriguez:2016vmx, Mapelli:2021taw}. For example, high mass can imply that a black hole may not be the direct result of stellar evolution and could have undergone previous mergers or accretion \citep{2017ApJ...840L..24F, Kimball:2019mfs}, while black hole spins misaligned from the orbital axis suggest formation other than from an isolated stellar binary.

Orbital eccentricity is the third axis, beyond mass and spin, that carries astrophysically relevant information. Gravitational-wave emission circularizes the binary orbit over time \citep{Peters_1964, 1963PhRv..131..435P}; therefore, long-lived, unperturbed binaries will not be eccentric by the time they reach detectable gravitational wave frequencies. Recently formed binaries, for example in dynamical encounters, may however not have enough time before merger to circularize \citep{Rodriguez:2017pec}. Binaries residing in the vicinity of supermassive black holes may also gain eccentricity via the exchange of orbital inclination and eccentricity, called the Kozai-Lidov mechanism \citep{1962_Kozai, 1962_Lidov}. Alternatively, binaries residing within the disks of active galactic nuclei can gain eccentricity both through interaction with the disk gas and due to close encounters with black holes and stars in the galactic center \citep{Tagawa:2020jnc, 2022Natur.603..237S}.

While the process of reconstruction of mass and spin from a gravitational wave signal is well established, determining orbital eccentricity has been lagging behind. One reason is the fact that eccentricity increases the binary parameter space, making it more difficult and computationally expensive to fully model it. In addition, due to binary circularization through gravitational wave emission, it had been anticipated that most binaries would not be eccentric, delaying the research and development of eccentric waveforms. 

Several recent works found signs of eccentricity for some of the detected black hole mergers. The most studied candidate has been GW190521 \citep{2020PhRvL.125j1102A}, which was found to have indications of high eccentricity by three independent studies \citep{GayathrieBBH,2020ApJ...903L...5R,2021arXiv210605575G} (although see \citealt{iglesias2023eccentricity, Gupte:2024jfe}). These three studies relied on different waveform models \citep{Healy:2022wdn, Cao:2017ndf, Ramos-Buades:2021adz, Chiaramello:2020ehz} and different analysis techniques that each had limitations, highlighting the need to comprehensively study the models and techniques involved. 

%Recently an eccentric binary merger {\it Task Force} assembled by the LIGO/Virgo/KAGRA collaborations \citep{2015CQGra..32g4001L, VIRGO:2014yos, KAGRA:2018plz} made recommendations on a list of studies that could establish the utility and accuracy of eccentric analyses \citep{eBBHtaskforcereport}. The goal of these recommendations is to develop and understand the tools to identify eccentric binary mergers and to accurately reconstruct their properties. 

In this work, we aim to address the following objectives: 
%recommendations of the LIGO/Virgo/KAGRA eccentric Task Force: 
\begin{enumerate}[align=left]
  \item Perform search sensitivity studies using available numerical relativity (NR) simulations.
  \item Perform sensitivity studies using available eccentric waveform models.
\end{enumerate}

These tests help establish the sensitivity of searches, which is necessary to observationally constrain astrophysical models predicting eccentric mergers. In addition, carrying out sensitivity studies for multiple waveform types helps determine the consistency of these waveforms, at least to the extent of search sensitivity.

The paper is organized as follows. In Section \ref{sec:Waveforms}, we describe the waveform simulations and models considered in this study.  In Section \ref{sec:comparison}, we present a direct comparison of a numerical relativity waveform catalog and a waveform model to directly test consistency across the parameter space. In Section \ref{sec:Search}, we describe the deployed search algorithm and injection strategy. In Sections \ref{sec:Results} and \ref{sec:Conclusions}, we discuss the results and conclusions of this work.

\section{Eccentric Waveforms}
\label{sec:Waveforms}

To determine how sensitive our searches are to gravitational wave signals from eccentric binaries, we used simulated waveforms that describe such systems. Significant progress has been made in the past decade in the development of eccentric binary waveforms using different approaches. In this paper, we utilize waveforms that were generated using two different methods that target different parts of the parameter space that we are interested in. This is necessary as each currently available waveform only covers a part of the parameter space, while this also enables us to compare the performance of the waveforms. 

\subsection{Numerical Relativity Simulations and Definitions of Eccentricity}
\label{sec:NR}

Numerical Relativity (NR) has made significant progress in accurately evolving the full Einstein field equations for binary black hole systems \citep{Mroue:2013xna, Chu:2015kft, SXS_Collab_2019, Jani:2016wkt, Healy:2017psd, Healy:2019jyf, Healy:2020vre}. More recently, there have been efforts in the NR community to include the effects of orbital eccentricity in these simulations. 

We adopted NR eBBH waveforms from the fourth RIT waveform catalog \citep{Healy:2022wdn} to cover the eccentricity range $0 \leq e < 0.9$. These simulations are relatively short and therefore target only high-mass systems with total binary mass $m_{\rm tot}\gtrsim 100M_\odot$. 

We additionally adopted NR eBBH waveforms from the SXS Collaboration  \citep{Hinder:2017sxy, SXS_Collab_2019}, to check for consistencies between the SXS and RIT NR waveform catalogs. These waveforms are longer and can cover lower masses ($m_{\rm tot}\gtrsim 70 \, M_\odot$) at fixed initial frequency, but are limited in eccentricity ($0 \leq e \leq 0.3$). 

A critical issue we took into consideration in this work is that the above two waveform families use different definitions of eccentricity. In \cite{O3_eBBH_Collab}, the SXS waveforms were employed with the eccentricity definition of \cite{2022PhRvD.106l4040R}, which is derived from the earlier eccentricity definition of \cite{Mora_eccentricity}. This is based on the orbital frequencies of the gravitational wave signal at closest and farthest approach. The modification introduced by \cite{2022PhRvD.106l4040R} ensures that the new eccentricity definition has the correct Newtonian limit while also eliminating the coordinate-dependence of the earlier definition. Another notable advantage of this definition is that it enables us to calculate the eccentricity at any point of the binary evolution as long as there are multiple orbital periods in the simulation \citep{Bonino:2022hkj}. However, in the cases when the binary orbit is highly eccentric and the two bodies directly plunge without notable inspiral cycles, this definition cannot be used. 
A remedy in this case is to use gauge-invariant combinations of energy and angular momentum, introduced in \citep{Albanesi:2023bgi,Carullo:2023kvj}.

The RIT waveform family employs a different, instantaneous eccentricity definition that can be used even if the simulated waveform duration is less than one orbital period \citep{GayathrieBBH, RIT_ecc_defn}. In this framework, eccentricity is defined at the apocenter to be $e=2\epsilon - \epsilon^{2}$, where $\epsilon=1-p_{\rm t}/p_{\rm t,qc}$. Here, $p_{\rm t}$ is the binary's tangential momentum, while $p_{\rm t,qc}$ is what the same binary's momentum would be at the same separation if the orbit was quasicircular. This definition, however, only enables the characterization of the eccentricity at the initial point of the simulated evolution. 

A full list of the NR waveforms used in this study can be found in Appendix {\ref{sec:appendix:NR}}.

\subsection{Effective One Body Formalism Waveforms}
\label{se:TEOB}

In the low-mass parameter space, we require semi-analytical models to generate eBBH waveforms, since it would be computationally very expensive to use NR simulations in this regime due to the required duration of the simulation to cover the full detectable frequency band. 

For this purpose we adopted the gravitational waveform model based on the effective-one-body (EOB) formalism, called TEOBResumS \citep{Damour:2014sva, Chiaramello:2020ehz, Nagar:2021gss}. In this model, the part of the gravitational waveform up to merger is generated by resumming post-Newtonian results. The final portion of the waveform is constructed using quasi-circular, NR-informed merger and ringdown \citep{Damour:2014yha, Nagar:2020pcj, Riemenschneider:2021ppj}. This latter choice is due to its simplicity and the limited availability of eccentric NR waveforms covering the required parameter space. 

The eccentric TEOBResumS model, dubbed as TEOBResumS-Dali, generates stable waveforms up to an eccentricity of about $e\sim 0.9$. However, the waveforms have only been validated with mildly eccentric NR waveforms from the SXS collaboration up to an eccentricity of $\sim 0.3$ \citep{Chiaramello:2020ehz, Nagar:2021gss, Bonino:2024xrv}, and up to moderate eccentricities in \cite{2021arXiv210605575G, Albanesi:2024xus}, making the reliability of the waveforms at higher eccentricities uncertain. Therefore, our results below also represent a test of these waveforms at higher eccentricities in comparison to NR waveforms. 

In \cite{Albanesi:2024xus}, the TEOBResumS-Dali model was also validated against waveforms representing initially unbound, comparable-mass binary black hole systems, which result in either dynamical captures or scatterings. \cite{Andrade:2023trh} generated a new suite of NR waveforms representing dynamical captures  using the \texttt{EinsteinToolkit} Software \citep{Loffler:2011ay}, and used these to validate the TEOBResumS-Dali model.

Future work will involve consistency checks of the TEOBResumS model incorporating eccentricity-informed merger and ringdown amplitudes computed in \cite{Carullo:2023kvj,Carullo:2024smg}.
This modification will improve the model's accuracy for high eccentricities.

\subsection{Prescription for Initial Conditions of Waveforms}

Since eccentricity is a time-dependent quantity, and given the multiple methods to compute it, it is important to ensure for our comparison that eccentricity definitions are consistent between waveforms. There have been various approaches to define eccentricity of binary compact object orbits, a summary of which can be found in \cite{Loutrel_2018}. To ensure that the waveforms from each approach are generated with the same parameters, we specified the initial conditions as physical phase-space variables rather than gravitational-wave parameters $f_\mathrm{low}$ and eccentricity $e$. The eccentricity at any point in time can then be calculated \textit{a posteriori} using properties of the generated waveform. For the actual eccentricity calculation, we adopted the prescription followed in \cite{O3_eBBH_Collab}, based on \cite{Mora_eccentricity, 2022PhRvD.106l4040R, Shaikh:2023ypz}. 

In the cases where we do not have sufficient gravitational-wave cycles to use this prescription, we use a definition of eccentricity which can be derived directly from the phase-space variables \citep{GayathrieBBH, RIT_ecc_defn}. We additionally also verified that the two different approaches give similar estimates of eccentricity (within $5 \%$) in the parameter space where both definitions are employable.

Eccentric waveforms for the TEOBResumS model are generated by specifying the initial conditions of the system using mass-reduced phase-space variables $(r_0, E_0, p_{0}^{\phi})$ along with other source properties like the total mass, $M \equiv m_1 + m_2$ and mass ratio $q \equiv m_2/m_1$. The EOB variables are related to physical variables as: (i) $r_0=R/M$ derived from $R$, the relative separation between the two bodies, (ii) $E_0= E_i / M$ derived from $E_i$, the initial energy of the system and (iii) $p_{0}^{\phi} = P_{0}^{\phi}/(\mu M)$ derived from $P_{0}^{\phi}$, the initial angular momentum of the binary, where $\mu \equiv m_1 m_2 / M$. We optimize $(r_0, E_0, p_{0}^{\phi})$ about the values from the metadata of the chosen NR waveform set to ensure that both waveform models are generated with the same initial conditions. The system evolution begins at the apastron.

This alternative method of defining the initial conditions for eccentric waveforms using phase-space variables instead of the traditional eccentricity and initial frequency allowed us to avoid discrepancies arising from the varying definitions of eccentricity between models, while also enabling us to explore the high-eccentricity parameter space. Furthermore, this method ensures that we are able to generate stable waveforms with the TEOBResumS model for the entire eccentricity space.

\section{Waveform Model Consistency by Direct Comparison}
\label{sec:comparison}

%We evaluated the consistency between our adopted waveform models, and more specifically, the parts of the parameter space where this consistency applies. 

As an initial check for consistency, we compared the signal strengths of the waveforms obtained with our two NR models for a range of eccentricities. Signal strength is quantified using the signal-to-noise ratio (SNR) which is calculated as:
\begin{equation}
\mathrm{SNR} = \int_{0}^{\infty} df \frac{4 |\tilde{h}(f)|^2}{S_{n}(f)}
\end{equation}
$\tilde{h}(f)$ represents the frequency-domain waveform obtained through a Fourier transform of the corresponding time-domain waveform $h(t)$. $S_{n}(f)$ represents the one-sided power spectral density (PSD) \citep{Cutler:1994ys} of the detector noise. For the SNR computations in this paper, we used the detector noise from the third observing run of LIGO/Virgo/KAGRA.

\begin{figure*}
\center
\includegraphics[scale=0.47]{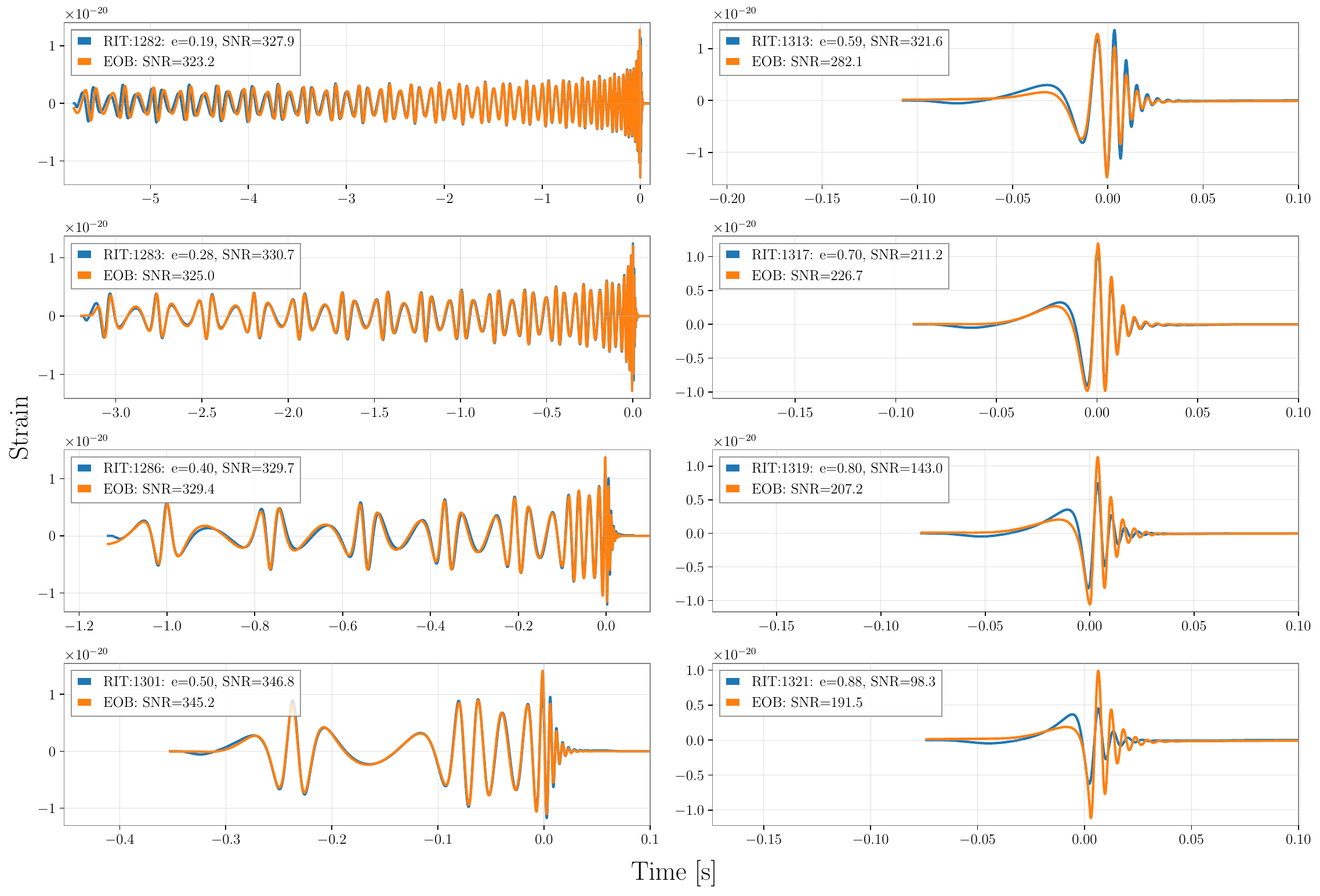}
\caption{Comparison of time-domain waveforms for the two different waveform approximants (indicated in the legend) for equal mass binary systems with total mass = $100 \, M_{\odot}$. The simulations start at an orbital separation that translates to an orbital frequency (at apastron) of $\mathrm{f_{low}=5\,\mbox{Hz}}$. The eccentricity values indicated in the legend are defined at the same $\mathrm{f_{low}}$.}
\label{fig:waveform_consistency}
\end{figure*}

Figure \ref{fig:waveform_consistency} shows the time-domain waveforms and corresponding SNRs for select waveforms from both RIT and TEOBResumS models. We see that the SNRs are in good agreement with each other up to an eccentricity of $\sim 0.7$. For low to moderate eccentricities, the quasi-circular merger of the TEOBResumS model does very well to accurately describe the process because the binary has sufficient cycles to circularize before merger. 
It is unclear whether residual differences are due to waveform inaccuracies or intrinsic NR error, given the availability of single resolution NR waveforms only.
For higher eccentricities, the distance of closest approach is comparable in magnitude to the last stable orbit (LSO) and therefore the binary merges quickly. As shown in Figure \ref{fig:waveform_consistency}, a quasi-circular merger-ringdown is not sufficient to accurately represent the final stages of the binary coalescence \citep{Carullo:2023kvj}, as the binary does not have adequate time to circularize. The discrepancy in the SNRs for the two waveforms at high eccentricities is therefore justifiable. In section \ref{sec:results_sensitivity}, we will describe the results of consistency checks that we performed within the framework of our search algorithm.

\section{Search Algorithm and Injections}
\label{sec:Search}

A key question for detecting eBBH coalescenses is the rate density of their occurrence in different astrophysical formation scenarios. Computing the rate density based on measurements requires the quantification of how sensitive we are to such sources. We computed this sensitivity as a function of source parameters by injecting simulated eBBH waveforms with various source parameters into the data obtained from the third observing run of LIGO/Virgo/KAGRA (O3) \citep{2020PhRvD.102f2003B}, and detect them with our search algorithm of choice. Using the ratio of recovered to injected signals, we estimated our sensitivity as a function of total binary mass $M$ in the source frame, mass ratio $q$ and eccentricity $e$. 

We additionally used the estimated sensitivity to check for consistency between the two waveform models.

\begin{figure}
\centering
\captionsetup[subfigure]{skip=5pt}  
    \begin{subfigure}{0.48\textwidth}
        \centering
        \includegraphics[scale=0.28]{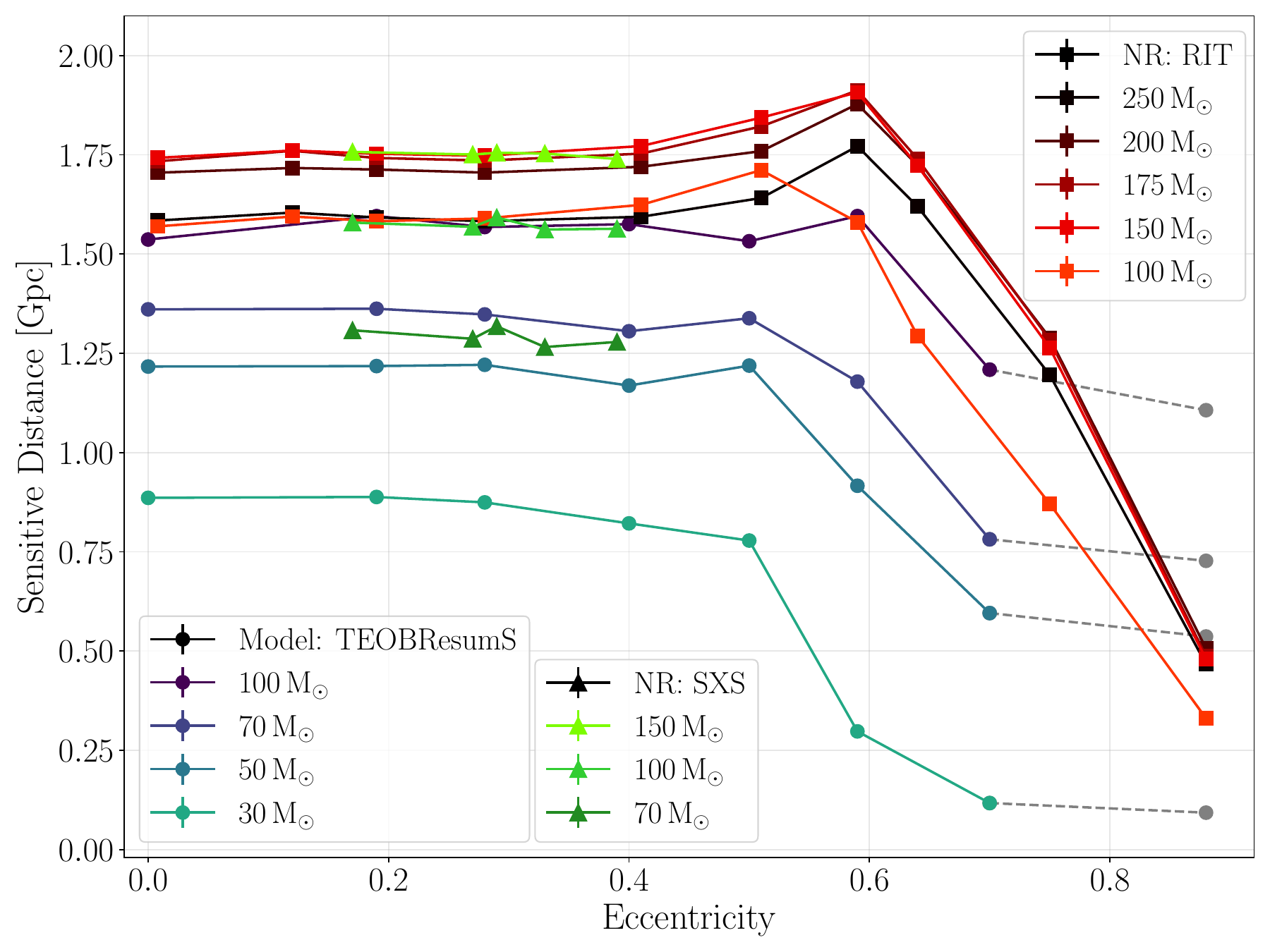}
        \caption{}
        \label{fig:sensitive_distance_q1}
    \end{subfigure}
    \vspace{0.1cm}
    \begin{subfigure}{0.48\textwidth}
        \centering
        \includegraphics[scale=0.28]{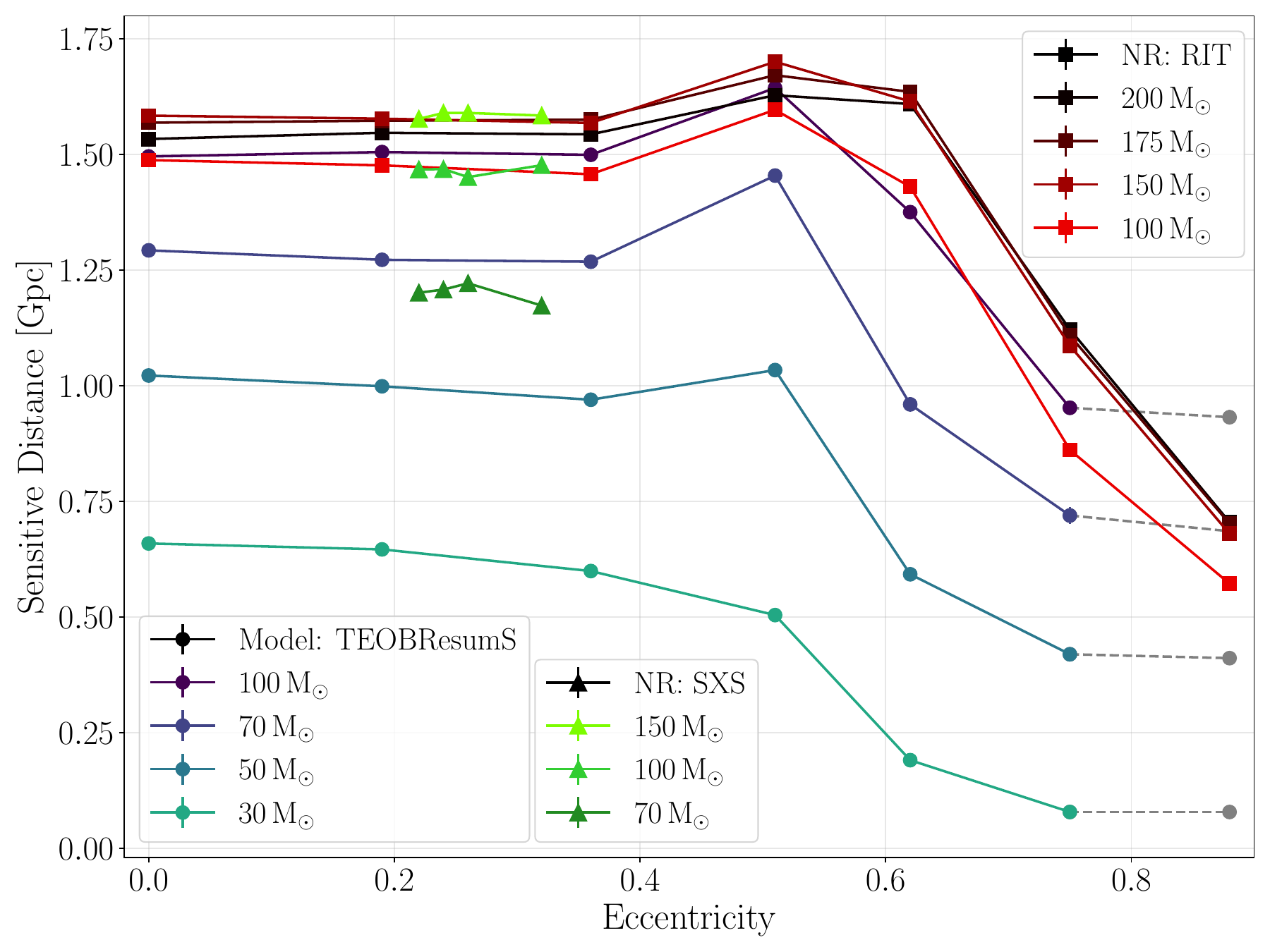}
        \caption{}
        \label{fig:sensitive_distance_q2}
    \end{subfigure} 
    \caption{Sensitive distance as a function of eccentricity for NR and TEOBResumS waveforms for masses ranging from $30\,M_{\odot}$ to $250\,M_{\odot}$ for an IFAR threshold of 10 yr. The subfigures \ref{fig:sensitive_distance_q1} and \ref{fig:sensitive_distance_q2} are for mass ratios $q=1.0$ and $q=0.5$ respectively. Eccentricities are defined at an orbital separation that corresponds to gravitational-wave frequency of $\mathrm{f_{low}=5\,\mbox{Hz}}$ for a system with total mass $100\,M_{\odot}$. In both cases, when we see significant deviations between sensitive distance with TEOBResumS and NR injections, we denote these instances with grey dashed lines.}
    \label{fig:sensitive_distance}
\end{figure}

\subsection{Search Pipeline}
\label{sec:cWB}
For our search, we employed coherent WaveBurst (cWB; \citealt{Klimenko:2005xv, 2008CQGra..25k4029K,cWB_MaxL, 2016PhRvD..93d3007T}), a model-agnostic search pipeline that uses minimal assumptions about the signal waveform. No comprehensive eccentric waveform template bank currently exists (template based searches that use quasi-circular templates have some, albeit reduced, sensitivity to eccentric mergers \citep{2010PhRvD..81b4007B, Gadre:2024ndy}).%, making template-based searches less sensitive to eBBH sources, especially for high eccentricities.

The cWB algorithm looks for excess power in time-frequency representations of detector data using appropriate clustering algorithms. Clusters which have energy above the expected detector noise fluctuations are categorized as events by the cWB algorithm. For each event, cWB constructs summary statistics that describe different properties of the events, for example, the duration, central frequency of the signal, correlation of the signal across the detectors, etc. Thresholds are placed on these summary statistics to better distinguish between true gravitational-wave events and noise fluctuations that can mimic gravitational-wave signals. This technique also works to increase the significance of candidate events. The cWB search performs well for chirp masses ${\cal M}_{\rm c} > 30$\,M$_\odot$, and its sensitivity remains robust to changes in eccentricity for $e < 0.6$ \citep{Gadre:2024ndy}. For our analysis, we used thresholds that maximized sensitivity to signals from eBBH coalescences. These optimized eBBH thresholds were introduced in \cite{O3_eBBH_Collab}.

\subsection{Injections}
\label{sec:sensitivity_analysis}
We injected simulated eBBH signals into detector data from the first half of the third observing run of LIGO/Virgo/KAGRA (O3a). The total mass parameter space, $M \in [30\,M_{\odot}, 250\,M_{\odot}]$ was covered using the different waveform models described in Section \ref{sec:Waveforms}. For the TEOBResumS waveform model we generated signals within the total mass range $M \in [30\,M_{\odot}, 100\,M_{\odot}]$. For the SXS NR waveforms, we generated injections within the total mass range $M \in [70\,M_{\odot}, 200\,M_{\odot}]$. For the RIT NR waveforms we generated injections in the $M \in [100\,M_{\odot}, 250\,M_{\odot}]$ total mass range. For each eccentricity, the waveforms were generated using the $(r_0, E_0, p_{0}^{\phi})$ values obtained from the metadata of NR waveforms corresponding to that eccentricity. 

The injected signals for each fixed source parameters of $(M, e, q)$ were uniformly distributed in sky location ($\theta, \phi$) and inclination $\iota$. Injections were performed roughly every 100\,s throughout O3a data. The signals were distributed uniformly in comoving volume up to a maximum redshift that was specific to each $(M, e, q)$, ensuring that the number of injections well beyond detection range are minimized. Sensitive volume-time (VT) and sensitive distance (defined in Eq. (5) and (6) of \cite{Chandra:2020ccy}) were calculated using the fraction of recovered to injected signals, where a recovered signal is one that has an inverse false alarm rate (IFAR) $\geq 10$\,yr.

\section{Results}
\label{sec:Results}

\subsection{Search Sensitivity}
\label{sec:results_sensitivity}

Figure \ref{fig:sensitive_distance} shows search sensitivity results for the entire mass space, separately for mass ratios $q=1.0$ and $q=0.5$. For both mass ratios under consideration, we find that there is strong agreement between the sensitivities obtained using the RIT and TEOBResumS waveforms up to moderately high eccentricities. Beyond $e_0=0.7$ however, the mismatches increase significantly. As seen in Figure \ref{fig:waveform_consistency}, beyond this eccentricity, the waveforms essentially only comprise of the merger. Since the TEOBResumS waveforms employed in this study are calibrated to quasi-circular NR waveforms for the merger phase, they do not accurately capture the physics of the merger-ringdown phase of the eccentric waveform. We see this trend in lower masses as well, where we would expect sensitivities to drop for highly eccentric systems. Instead we see a saturation in sensitivity for $e>0.7$ across the mass bins and mass ratios that were examined. Therefore, we conclude that within the context of our unmodeled search algorithm, current TEOBResumS waveforms \citep{Damour:2014sva, Chiaramello:2020ehz, Nagar:2021gss} are valid up to initial eccentricities of 0.7, defined at initial separations that correspond to $f_\mathrm{low} = 5$\,Hz for a binary with total source mass $M = 100 M_{\odot}$.

We further find strong agreement between the sensitivities obtained for the SXS and RIT NR waveforms for the same parameters. This demonstrates that, at least to the extent of sensitivity, the two NR waveform families are consistent with each other.

\subsection{Astrophysical Implications}
\label{sec:results_astrophysical_implications}

In addition to gaining an understanding of the waveform systematics relating to the TEOBResumS model, a benefit of characterizing the sensitivity as a function of $(M, e, q)$ is that it allows us to calculate sensitive volume-time (henceforth referred to as just $VT$) for any astrophysical population for which we know the distribution in $(M, e, q)$. From the cWB sensitivity study, we have $VT$ as a function of $M$, $e$ and $q$ for discrete points in this three-dimensional space. We interpolate between the available points to obtain a continuous $VT (M, e, q)$. With this function and available distributions of eccentric binary populations from literature ($f(M, e, q)$), we can compute an effective $VT$ for each population which is given by 

\begin{multline}
\left\langle VT \right\rangle = \int_{30 M_{\odot}}^{250 M_{\odot}}\ \int_{0}^{e_\mathrm{high}} \int_{0.5}^{1.0} VT(M, e, q) \, f(M,e,q)\, \\dM\,de\,dq
\end{multline}

Here, $e_\mathrm{high}$ is the upper limit in eccentricity for which the waveforms have been shown to be valid. The eccentricities must be defined at a fixed frequency, as the available information on eccentricity distributions for different astrophysical populations corresponds to a fixed gravitational-wave emission frequency, defined at apastron. We used the post-Newtonian (PN) expressions derived in \cite{Tanay:2016zog} to extrapolate our eccentricities to a gravitational-wave emission frequency of $15$\,Hz. We however observe that for $e_0 > 0.6$, this prescription yields unphysical results. 
In this study, we therefore limited the $VT$ computations to only include results from waveforms that have initial eccentricity, $e_0 \leq 0.6$ for all mass bins. 

Once we obtained $VT$ for each population under consideration, we place upper limits to merger rates assuming that all the gravitational-wave candidates that have been identified so far are consistent with non-eccentric binaries. Our results are shown in Table \ref{tab:population}.

\begin{table}[h]
\begin{center}
\begin{tabular}{ c  c  c  c }

\hline
 \multirow{2}*{$p(M)$} & \multirow{2}*{$p(q)$}  & \multirow{2}*{$p(e)$} & $\mbox{VT}$ \\
 & & & [Gpc$^{3}$yr] \\
\hline
 GWTC-3 & GWTC-3 & $2(1-e)$ & $1.49$ \\

 GWTC-3 & GWTC-3 & uniform & $1.49$ \\

 $M^{-2.3}$ & uniform & $2(1-e)$ & $2.04$ \\

 $M^{-2.3}$ & uniform & uniform & $2.02$ \\

 AGN & AGN & $2(1-e)$ & $2.61$ \\

 AGN & AGN & uniform & $2.65$ \\

 %DSC & DSC & DSC & \ToDo{-} \\
\hline

\end{tabular}
\caption{Total volume--time covered by our search for various source total mass, mass ratio, and eccentricity distributions.}
\label{tab:population}
\end{center}
\end{table}

\section{Conclusions}
\label{sec:Conclusions}

We carried out a sensitivity and consistency study for three eccentric binary merger waveform families, including the SXS and RIT NR waveforms and the TEOBResumS semi-analytical waveforms. We determined the sensitivity of the LIGO detectors during their O3a observing run to eccentric waveforms as a function of mass, mass ratio and eccentricity (assuming zero spin). Our conclusions are summarized as follows:
\begin{itemize}[align=left]
\item The current version of the TEOBResumS-Dali waveform model, which uses a quasi-circular merger-ringdown prescription, is valid for GW searches up to $e\sim0.7$ for the initial frequencies considered in this work.
\item We found close to equal sensitivities for the SXS, RIT and TEOBResumS ($e<0.7$) waveforms for the mass and eccentricity ranges at which we were able to compare them to each other (see Fig. \ref{fig:sensitive_distance}). This supports the utility of each of these waveforms to determine search sensitivity, and is encouraging for their potential utility in parameter estimation.
\item With the obtained search sensitivity we found the VT probed during the O3 observing run of LIGO/Virgo/KAGRA, generally around $1.5-2.5$\,Gpc$^{3}$yr with some model dependency, as seen in Table \ref{tab:population}. Assuming non-detection of eccentric events, this corresponds to an upper limit of $\sim 1-1.5$\,Gpc$^{-3}$ yr$^{-1}$ on eBBH merger rates at 90\% confidence level \citep{Biswas:2007ni, LIGOScientific:2016hpm, O3_IMBH}. This is consistent with theoretical merger rate predictions in literature \citep{Rodriguez:2018pss, 2021ApJ...921L..43Z}.
\end{itemize}

\begin{acknowledgments}
The authors gratefully acknowledge Rossella Gamba for valuable discussions and assistance with the TEOBResumS code. The authors thank Alessandro Nagar for providing useful feedback on the manuscript. The authors additionally thank the SXS and RIT collaborations for maintaining their public waveform databases. The authors acknowledge support by the National Science Foundation (NSF) grant PHY 2309024.
This material is based upon work supported by NSF's LIGO Laboratory which is a major facility fully funded by the National Science Foundation. The authors acknowledge the computational resources that aided the completion of this project, provided by LIGO-Laboratory and supported by the NSF Grants No..PHY-0757058 and No.PHY-0823459.
G.V. acknowledge the support of NSF under grant PHY-2207728.
S.K. and T.M. acknowledge the support of NSF under grants PHY 2110060 and PHY-2409372.
G.C. acknowledges funding from the European Union’s Horizon 2020 research and innovation program under the Marie Sklodowska-Curie grant agreement No. 847523 ‘INTERACTIONS’, and support from the Villum Investigator program by the VILLUM Foundation (grant no. VIL37766) and the DNRF Chair program (grant no. DNRF162) by the Danish National Research Foundation.
This project has received funding from the European Union's Horizon 2020 research and innovation programme under the Marie Sklodowska-Curie grant agreement No 101131233.
M.S. acknowledges Polish National Science Centre Grant No. UMO-2023/49/B/ST9/02777 and the Polish National Agency for Academic Exchange within Polish Returns Programme Grant No. BPN/PPO/2023/1/00019.
J.L.
C.O.L and J.H. gratefully acknowledge NSF for financial support from Grant
PHY-2207920 and used the ACCESS TGPHY060027N and project PHY20007 at Frontera (TACC) allocations to perform the full numerical simulations of binary black holes.
The version of TEOBResumS used in this work is available publicly at \url{https://bitbucket.org/teobresums/teobresums/src/Dali/} (\texttt{commit: 611dde3}).

\end{acknowledgments}
%%%%%%%%%%%%%%%%% APPENDICES %%%%%%%%%%%%%%%%%%%%%

\appendix

\section{List of NR Waveforms}
\label{sec:appendix:NR}

 Table \ref{tab:NR} lists the numerical relativity simulations used in this study. The eccentrcities cited here are defined at an orbital separation that corresponds to a gravitational-wave frequency of $5$\,Hz (defined at apastron) for a binary system with total source-frame mass $100\,M_{\odot}$. Since the SXS simulations start at a higher frequency, we extrapolated the eccentricities to $5$\,Hz using the post-Newtonian expressions derived in \cite{Tanay:2016zog}.

\begin{table}[h]
\begin{center}
\begin{tabular}{ c  c  c  }
\hline
 $q$ & $e$ & Waveform ID\\
\hline
 $0.5$ & $0.22$ & SXS:BBH:1365\\

 $0.5$ & $0.24$ & SXS:BBH:1366\\

 $0.5$ & $0.26$ & SXS:BBH:1369\\

 $0.5$ & $0.33$ & SXS:BBH:1370\\

 $1.0$ & $0.09$ & SXS:BBH:1355\\

 $1.0$ & $0.14$ & SXS:BBH:1357\\

 $1.0$ & $0.25$ & SXS:BBH:1361\\

 $1.0$ & $0.23$ & SXS:BBH:1362\\

 $1.0$ & $0.37$ & SXS:BBH:1363\\

 $0.5$ & $0.00$ & RIT:eBBH:1200\\

 $0.5$ & $0.19$ & RIT:eBBH:1422\\

 $0.5$ & $0.36$ & RIT:eBBH:1423\\

 $0.5$ & $0.51$ & RIT:eBBH:1426\\

 $0.5$ & $0.62$ & RIT:eBBH:1429\\

 $0.5$ & $0.75$ & RIT:eBBH:1433\\

 $0.5$ & $0.88$ & RIT:eBBH:1436\\

 $1.0$ & $0.00$ & RIT:eBBH:1090\\

 $1.0$ & $0.19$ & RIT:eBBH:1282\\

 $1.0$ & $0.28$ & RIT:eBBH:1283\\

 $1.0$ & $0.40$ & RIT:eBBH:1286\\

 $1.0$ & $0.50$ & RIT:eBBH:1301\\

 $1.0$ & $0.59$ & RIT:eBBH:1313\\

 $1.0$ & $0.70$ & RIT:eBBH:1317\\

 $1.0$ & $0.88$ & RIT:eBBH:1321\\

\hline 
\end{tabular}
\caption{Parameters of the numerical relativity simulations adopted from the SXS binary black hole simulations Catalog \citep{Hinder:2017sxy, SXS_Collab_2019} and the 4$^{\rm th}$ RIT  catalog \citep{Healy:2022wdn}. Columns show the binary's mass ratio, and initial eccentricity. The spin amplitudes $S_1$ and $S_2$ of the two black holes in the binary are 0 for all simulations in this table.}
\label{tab:NR}
\end{center}
\end{table}

%\section{EOB Dynamics}
%Plot showing EOB dynamics for e>=0.6.

%%%%%%%%%%%%%%%%%%%%%%%%%%%%%%%%%%%%%%%%%%%%%%%%%%

\newpage
\bibliography{References.bib} 

%%%%%%%%%%%%%%%%%%%%%%%%%%%%%%%%%%%%%%%%%%%%%%%%%%

\end{document}